\documentclass[preprint,12pt]{elsarticle}
\usepackage[utf8]{inputenc}

\usepackage{amssymb}
\usepackage{amsmath}
\usepackage{multirow}
\usepackage{xcolor}

\journal{Journal of Subatomic Particles and Cosmology}

\newcommand{\clr}[1]{\textcolor{black}{#1}}

\begin{document}

\begin{frontmatter}



\title{Machine Learning for Extrapolating No-Core Shell Model Results to Infinite Basis}


%\author{} %% Author name

%\affiliation{organization={},%Department and Organization

\author[PNU]{R. E. Sharypov}
\ead{2017104939@togudv.ru}

\author[PNU]{A. I. Mazur}
\ead{amazur.pnu.khb@mail.ru}
%
%\author[PNU]{R. E. Sharypov}
%\ead{2017104939@togudv.ru}

\author[MSU]{A.~M.~Shirokov}
\ead{shirokov@nucl-th.sinp.msu.ru}

\affiliation[PNU]{organization={Pacific National University},%Department and Organization
            addressline={136 Tihookeanskaya street}, 
            city={Khabarovsk},
            postcode={680035}, 
%            state={Khabarovsk Krai},
            country={Russia}}

\affiliation[MSU]{organization={Skobeltsyn Institute of Nuclear Physics, Lomonosov Moscow State University},%Department and Organization
            addressline={Leninskie Gory 1/2}, 
            city={Moscow},
            postcode={119991}, %119234, ГСП-1, Москва, Ленинские горы, д. 1, стр. 2
            %state={},
            country={Russia}}

%% Abstract
\begin{abstract}
%{\color{red}  We explore the application of 
We utilize the machine learning %techniques 
to extrapolate to the  infinite model space 
the no-core shell model (NCSM) results for the energies and rms radii of the $^{6}$He ground state
and $^{6}$Li lowest states. %variational calculations to the  infinite model space. %s. The the no-core shell model  is a powerful \textit{ab initio} method for nuclear structure calculations, yet its computational cost grows exponentially with model space size, limiting practical calculations to relatively light nuclei. 
%Traditional extrapolation techniques lack rigorous theoretical justification, motivating the use of machine learning as an alternative. 
%We employ an ensemble of artificial neural networks trained on no-core shell model results for $^6$He and $^6$Li nuclei using the realistic Daejeon16 nucleon-nucleon interaction. 
%The proposed machine learning-based method improves the robustness and accuracy of extrapolated results by optimizing data preprocessing, training procedures, and selection criteria. 
The extrapolated % predicted ground-state 
energies and rms %root-mean-square 
radii converge as  the NCSM results from larger model spaces are included in the 
training dataset for ensemble of artificial neural networks thus enabling an accurate
predictions for these observables.
%exhibit strong convergence properties, and they are consistent with experimental data.
%This work demonstrates the potential of machine learning for advancing nuclear many-body theory by providing reliable extrapolations in computationally challenging scenarios.}
\end{abstract}

%\item The results obtained in this study highlight the effectiveness of the ensemble neural network approach for extrapolating No-Core Shell Model calculations to infinitely large basis spaces. The proposed method demonstrates a systematic improvement over previously reported machine learning-based extrapolations \cite{Negoita2019}, particularly in terms of statistical robustness and uncertainty quantification.
%% Keywords

\end{frontmatter}



\section{Introduction}
\label{Introduction}
%% Labels are used to cross-reference an item using \ref command.

%\section*{Machine Learning Methods in Nuclear Physics: Applications to Extrapolation in \textit{Ab Initio} Calculations}

Machine learning methods are %being 
increasingly utilized in both theoretical and experimental studies of atomic nuclei~\cite{Boehnlein2022}. In this paper, we continue exploring the %to explore the potential of 
machine learning techniques for extrapolating results of variational calculations of nuclear 
observables %properties 
to infinite model spaces.  

The no-core shell model (NCSM)~\cite{Barrett2013} is currently one of the most promising \textit{ab initio} approaches to %for 
theoretical investigations of nuclei. %nuclear properties. 
In the
NCSM, all nucleons are %treated as 
spectroscopically active, and their motion is governed exclusively by the chosen model of realistic nucleon-nucleon ($NN$) and, if needed, three-nucleon ($3N$) interactions, without
any phenomenological assumption. %any additional approximations. 
The NCSM basis states %in NCSM 
are constructed as Slater determinants of single-particle harmonic oscillator wave functions. The %primary parameters of 
NCSM parameters are the oscillator energy $\hbar\Omega$ and the size of the model space, characterized by the maximum number of excitation quanta $N_{\text{max}}$
 allowed for in the calculation. The NCSM basis functions for a nucleus with $A$ nucleons 
% in NCSM 
include all possible combinations of nucleons with  total oscillator quanta
%a %excitation quantum number 
ranging from %$N_{\text{min}}\hbar\Omega$ to $(N_{\text{min}} + N_{\text{max}})\hbar\Omega$,
$N_{\text{min}}$ to $N_{\text{min}} + N_{\text{max}}$, where $N_{\text{min}}$ is the %minimum 
% excitation
minimal oscillator quanta allowed for the nucleus as dictated by the Pauli exclusion principle.  

The NCSM results are exact in the limit $N_{\text{max}} \to \infty$. However, the %size
dimensionality of the many-body basis grows exponentially with increasing $N_{\text{max}}$, which restricts calculations with a reasonable precision, 
even on modern supercomputers, to light nuclei with mass numbers $A \lesssim 20$. 
%For the lightest nuclei, such as $^3\text{He}$, $^3\text{H}$, $^4\text{He}$, $^6\text{He}$, and $^6\text{Li}$, calculations are feasible up to $N_{\text{max}} \sim 20$. For nuclei in the middle of the $p$-shell, the maximum achievable $N_{\text{max}}$ is 12--14. 
Therefore, developing robust methods for extrapolating results to infinitely large model spaces remains a pressing challenge.  

Most extrapolation methods developed to date~\cite{Zhan2004,Maris2009,Coon2012,Maris2013,More2013, Kruse2013,Furnstahl2014, Shirokov2021} lack rigorous theoretical justification limiting their applicability. The most theoretically grounded approach~\cite{Shirokov2021} involves localizing the poles of the $S$-matrix for bound states.  

Machine learning offers a promising alternative for addressing this problem. This direction is also of independent interest %, 
as an application of the %applying 
machine learning %methods 
to extrapolation problems %extrapolate results of variational calculations 
requires the development of novel algorithms and theoretical approaches, opening new avenues for research.  

%\begin{comment}
%The first study of the %to explore the application of 
%machine learning extrapolation of %extrapolations of % methods for extrapolating 
%variational calculation results to the 
%infinite model space %spaces 
%was presented in Ref.~\cite{Negoita2019}. The proposed extrapolation method involves training an ensemble of artificial
%neural networks on NCSM results for a %calculations of 
%specific nuclear  observable %observables 
%performed with a set of %varying 
%values of the oscillator parameter $\hbar\Omega$ in model spaces up to some %$N_{\text{max}}^u$.
%maximal value of $N_{\max}=N_{\text{max}}^u$.
%\end{comment}

{%\color{violet}
The first study of the machine learning extrapolation of variational calculation results to the infinite model space was presented in Ref.~\cite{Negoita2019}.
The extrapolation method %introduced in this study
of Ref.~\cite{Negoita2019} (ISU) involves training an ensemble of artificial
neural networks to predict the target observable (e.\,g., energy $E$, rms point-proton radius~$r_p$, etc.) based on a set of calculations
of a nucleus of interest
 performed within the NCSM. Each neural network is a single-layer perceptron. The inputs are %features correspond to 
the NCSM parameters, %---
$N_{\max}$ and $\hbar\Omega$, %---
while the outputs are the NCSM results for %represent the computed values of 
the observable of interest obtained with these input parameters. 
The trained neural networks are used to predict the results at very large values of~$N_{\max}$
where the convergence of the NCSM calculations is supposed to be achieved.
Later a similar approach was %Subsequently, neural networks with a similar architecture were 
employed for the 
NCSM result extrapolations 
in Refs.~\cite{JiangPhysRevC.100.054326, VIDANA2023122625}.
}

{%\color{violet}
There is an alternative machine learning  approach TUDa to the extrapolation of NCSM calculations developed in 
Refs.~\cite{KNOLL2023137781,WolfgruberPhysRevC.110.014327}.
% which presents a different machine learning approach to the extrapolation of NCSM calculations. 
Instead of training neural networks for each particular nucleus, TUDa suggests an universal neural network ensemble trained on NCSM results 
for light nuclei where a complete convergence is achievable, using various $NN$ interactions. This universal ensemble is used for extrapolating the
NCSM results for heavier $p$-shell nuclei. A benchmarking of ISU and TUDa %extrapolations 
methods for the ground state energies and point-proton radii of Li
isotopes~\cite{Knol2025} demonstrated the consistency of the obtained results within the estimated uncertainties.
%: ``Instead of emulating the functional form of observables in terms of $\hbar\Omega$ and $N_{\max}$ the converged value is directly predicted from a set of calculations in small model
%spaces. This exploits the pattern recognition capabilities of artificial neural networks, which are designed to capture the observable-specific
%convergence patterns in few-body systems.''} {(\it ``text'' from \cite{Knol2025})
}

In this study, we continue the neural network extrapolation studies following the route suggested in the pioneering Ref.~\cite{Negoita2019} with 
modifications discussed below
 which we suggested in Refs.~\cite{Mazur2024,sharypov2024machine}. We present the extrapolated results 
%to investigate this direction. This paper presents predictions 
for the ground 
%of ground and excited 
state energies of the
$^6\text{He}$ and $^6\text{Li}$ nuclei and for
the lowest excited state energies of $^6\text{Li}$ 
as well as for the rms  point-proton $r_p$, point-neutron
$r_n$, and point-nucleon (matter) $r_m$ radii in 
%distribution radii of point protons $r_p$, point neutrons
%$r_n$, and point nucleons (matter) $r_m$ for 
these states, based on NCSM calculations conducted in model spaces with $N_{\text{max}} \leq 18$ using the realistic $NN$ %nucleon-nucleon 
interaction Daejeon16~\cite{SHIROKOV201687}.

The manuscript is organized as follows. %in the following way. 
A description of the method employed % in this work 
is presented in Section~\ref{Method}. The results of extrapolations of the considered observables
in $^6\text{He}$ and $^6\text{Li}$ %to large model spaces where the convergence is
%achieved, 
are presented
%the extrapolation to large model spaces of $^6\text{He}$ and $^6\text{Li}$ are shown 
and discussed in Section~\ref{Results}. Finally, a short summary and the conclusions %of this research 
are given in Section~\ref{Discussion}.

\section{Method}
\label{Method}

The outline of our method is that we employ an ensemble of randomly initialized 
artificial neural networks to analyze
results of the 
NCSM calculations and predict observables at very large $N_{\max}$ values where the convergence of the NCSM
calculations is supposed to be achieved.
%as $N_{\max}$ approaches infinity . 
The ensemble approach aims to improve the prediction %predictive 
accuracy and to assess its %the 
uncertainty.

As compared with the pioneering work~\cite{Negoita2019}, we proposed a different %We proposed a rather simple 
neural network topology and established 
strict criteria for selecting trained neural networks~\cite{Mazur2024,sharypov2024machine}. This eliminates %eliminated 
the need to split the data into training and testing sets, significantly simplifying the training process~\cite{Mazur2024}. Our %The 
modified approach, combined with rigorous statistical analysis of predictions from the selected networks, ensures statistical robustness and increases the %ncreased 
reliability of the extrapolated results.  

The proposed extrapolation method was tested on a benchmark problem of calculations of
the deuteron %: calculating the 
ground state energy supported by %of the deuteron using 
the realistic Nijmegen~II $NN$ potential~\cite{PhysRevC.49.2950}. 
%The exact ground-state energy for this problem, $E_{\text{exact}} = -2.224 \ \text{MeV}$, matches the experimental value. 
This problem %was chosen for several reasons. First, it 
exhibits a
slow convergence with
different behaviors of the results obtained with odd and even values of $N_{\text{max}}/2$ thus
posing additional challenges for extrapolation techniques, and %computational methods. Second, the observed parity-dependent behavior of $N_{\text{max}}/2$ 
provides a unique opportunity to verify the correctness and efficacy %efficiency 
of the developed algorithm. We note also that %Finally, 
the dependence on the %energy on the 
oscillator frequency $\hbar\Omega$ of the deuteron energy obtained with the Nijmegen~II,
demonstrates a
complex behavior characterized by %, including 
the formation of two 
local minima, making it particularly conclusive %suitable 
for testing machine learning extrapolation capabilities.  

\subsection{Neural network design}
\label{Method: Neural network design}

\begin{figure}[t]
    \centering
    \includegraphics[width=0.5\linewidth]{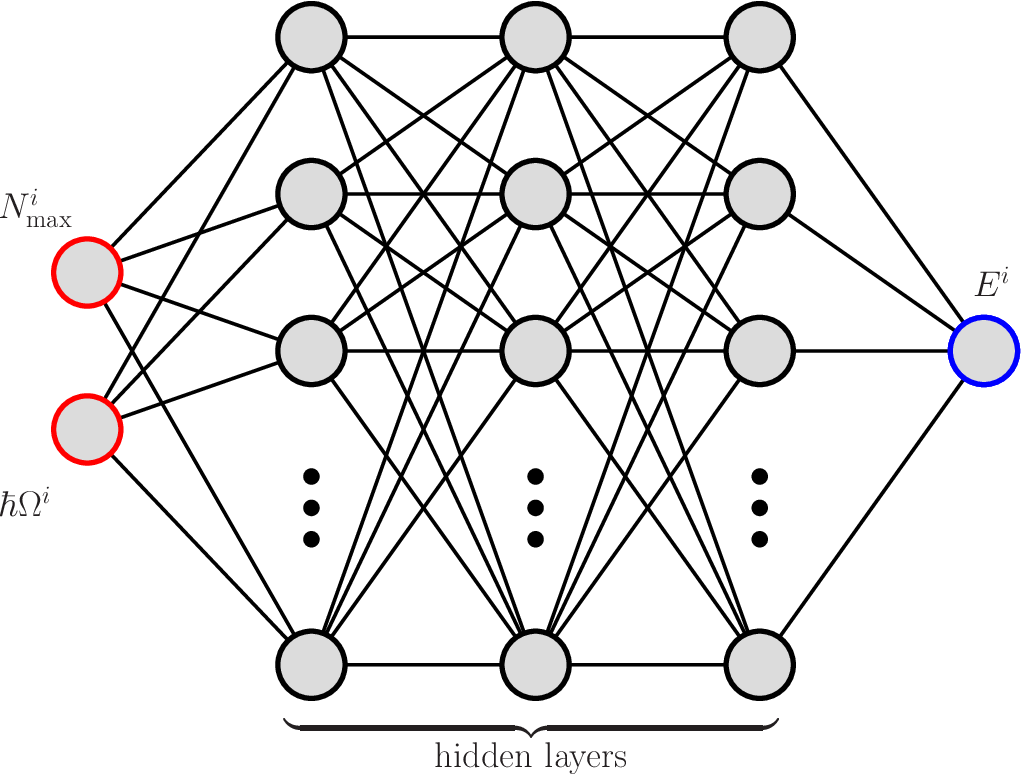}
    \caption{Neural network architecture}
    \label{Neural_network_acrhitecture}
\end{figure}

\paragraph{Network architecture}
\label{Method: Network architecture}
The proposed model consists of an ensemble of 1024 fully connected feedforward neural networks, each designed to handle the structured NCSM data. Each neural network %in the ensemble 
is multilayer perceptron with %and it follows an 
identical architectures %, 
comprising an input layer, three hidden layers, and an output layer as shown in Fig.~\ref{Neural_network_acrhitecture}.
The input layer consists of 2 neurons. One of the input layer neurons 
receives the  $N_{\max}$ value,
 another receives the~$\hbar\Omega$ value; 
 these values are transmitted to the neurons of the first hidden layer. 
Each neuron~$i$ of hidden %and output 
layers \clr{collects} %collect 
the signals~$y_{j}^{i}$ from all neurons of the previous layer,  sums them with its individual set of 
weights~$w_{j}^{i}$,  adds a bias~$b^{i}$ to obtain $x^{i}=\sum_{j} w_{j}^{i}y_{j}^{i}+b^{i}$ and uses the activation function~$f(x)$ to obtain the
signal~$f(x^{i})$ which is transmitted to all neurons of the next layer. The output layer does the same to produce the result~$f(x^{i})$ of the
calculation performed by the artificial neuron network. The weights~$w_{j}^{i}$ and biases~$b^{i}$ constitute a set of trainable network parameters.

The first hidden layer contains 10 neurons and employs the
identity activation function $f(x)=x$.
The second and the 
third hidden layers each contain 10 neurons and utilize the
sigmoid activation function ${f\left(x\right)=1/\big(1+\exp(-x)\big)}$. The output layer consists of a single neuron with the %and utilizes an 
identity activation function.
The random  %Weight 
initialization of weights and biases
follows the
default Glorot initialization method~\cite{Glorot_pmlr-v9-glorot10a}. 
%The machine learning
% to introduce difference of neural networks in the ensemble and enhance training stability and convergence. The model 
%is implemented using Keras \cite{chollet2015keras} with TensorFlow  \cite{tensorflow2015-whitepaper} as the backend.
{%\color{red} It introduces difference of neural networks in the ensemble and enhances training stability and convergence.
The machine learning model is implemented using Keras~\cite{chollet2015keras} with TensorFlow~\cite{tensorflow2015-whitepaper} as the backend. Programming code is available on GitHub \cite{code}.}

The chosen neural network architecture, particularly the use of the sigmoid activation function, leads to the saturation 
% {\color{red}of neuron connections} %of neuron connections 
when large $N_{\text{max}}$ values %of  
are provided as the input. % (this effect is also holds for $\hbar \Omega$). Consequently, 
As a result, in the limit $N_{\text{max}} \to \infty$, the neural network predictions
become independent of~$\hbar\Omega$.
% for the reasonable $\hbar \Omega$ range degenerate into line.

The use of the %a 
linear activation function in the output layer  %is crucial: it 
allows %the output layer 
to generate predictions over a wide range of~$\hbar\Omega$
without being constrained by the limiting values of the %an 
activation function; %. In this case, 
the saturation at large $N_{\max}$ % of neural network connections 
occurs in the network %its 
hidden layers.

%{\color{red}The use of an identity activation function in the output layer is crucial: it 
%allows the output layer 
%to generate predictions over a wide range of~$N_{\max}$ and $\hbar\Omega$ 
%without being constrained by the limiting values of the 
%activation function. Therefore 
%the saturation at large $N_{\max}$ % of neural network connections 
%occurs in the hidden layers of the network.}

%
%\begin{figure}[t]
%    \centering
%    \includegraphics[width=0.5\linewidth]{ANN_en.eps}
%    \caption{Neural network architecture}
%    \label{Neural_network_acrhitecture}
%\end{figure}
%

\paragraph{Data and Preprocessing}
\label{Method: Data and Preprocessing}
%{\color{red}The input features for the neural network are}
The neural network input includes %features for the neural networks include 
the NCSM parameters $N_{\text{max}}$ and $\hbar\Omega$, while the output corresponds %outputs correspond 
to the calculated observables (e.\,g., energy of a given state $E$, rms point-proton~$r_p$, point-neutron~$r_n$ or point-nucleon~$r_m$
%root-mean-square point proton, point neutron or point nucleon distribution 
radii).
The dataset %, 
composed of $N_{\text{max}}$, $\hbar\Omega$, $E$ (or $r_p$, or
$r_n$, or~$r_m)$ %, 
serves as the training set.
% {\color{red} after the following preprocessing}. % after the following preprocessing. %In Ref. \cite{Shirokov2021}, it 
It was
shown in Ref.~\cite{Shirokov2021}
that stable extrapolation results for energies
using the \mbox{SS-HORSE--NCSM} method are only achieved if, in each model
space,  the data lying to the right of the variational minimum in~%the 
$\hbar\Omega$ %in each model space 
are considered. Based on
our experience gained in numerical calculations, we
conclude that this selection of data also significantly
improves extrapolations %predictions of the method of extrapolation
utilizing machine learning.
The %Also the 
right cutoff value of $\hbar\Omega = 40$~MeV is applied both for radii and energies %energy 
while the 
left cutoff value for radii we %is 
set to $\hbar\Omega = 12.5$~MeV.
We use for the training the NCSM results obtained with a set of
different~$\hbar\Omega$ values in the model spaces $N_{\max}=4$,
$N_{\max}=6$, ...\,, $N_{\max}=N_{\max}^{u}$ and increase $N_{\max}^{u}$ to examine the convergence of the obtained predictions.
Prior to training, the data of each
type in the training set is normalized 
%the input features undergo normalization 
to a standard range of $\left[0,1\right]$.
% {\color{red} to improve performance of proposed approach on different extrapolation problems}. %to improve model performance on the different extrapolation problems.

\paragraph{Training process}
\label{Method: Training process}
Each network %model 
is trained using the Adam optimization algorithm \cite{Adam}, which provides internal adaptive learning rates and efficient gradient-based optimization.
Additionally, an external triangular cyclical learning rate schedule \cite{smith2017cyclical} is employed to dynamically adjust the learning rate during training, improving convergence.
This schedule is implemented with TensorFlow Addons \cite{TensorFlowAddons} library. 
The learning rate cycles between the
base value of $10^{-4}$ and the %cycle 
amplitude value which is equal to %of 
$10^{-2}$ at the beginning of the training and %, which 
decreases during the training. 
%Thus, this
This forms a decaying sawtooth pattern of the learning rate with %: 
large values providing a %provide 
rough convergence and smaller values providing the %, while small values provide 
fine tuning of the neural network.
%{\color{red} Such a method of selecting the learning rate schedule increases the likelihood that different neural networks within the ensemble will converge to substantially different local minima on the loss function manifold.}
%Such a method of selecting the learning rate schedule increases the likelihood that different neural networks within the ensemble will converge to substantially different local minima on the loss function manifold.

The batch size is fixed at the length of training set to balance computational efficiency and convergence speed. The loss function is defined as the 
mean-square deviation 
%{\color{red} mean-square error} %mean squared error for regression problem, 
ensuring minimization of the discrepancy between predicted and actual values. The training process is conducted over $10^6$ epochs. Additional details of the training can be found in Refs.~\cite{Mazur2024,sharypov2024machine}.

\begin{figure}[t]
    \centering    
    \includegraphics[width=0.48\linewidth]{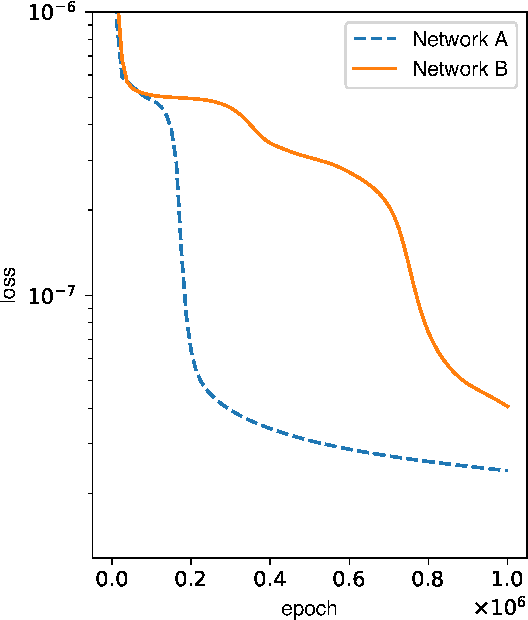} \hfill
    \includegraphics[width=0.48\linewidth]{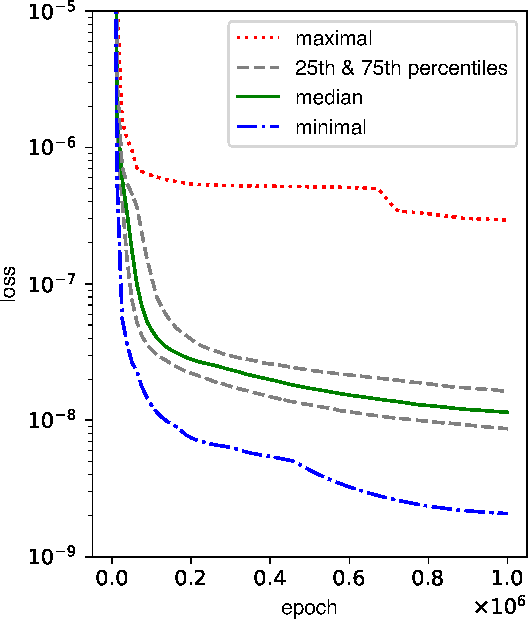}
    
    \caption{Evolutions of loss functions  %value 
during training on the $^{6}$Li %\(^6\text{Li} (3^+,0)\) energy data 
ground state energy datasets with \(N_{\max}^u=18\). Left panel: evolutions of two
different networks during the training; right panel: median loss function of the network ensemble  (green solid curve),  minimal (blue dash-dotted curve)
and  maximal (red dotted curve) loss function values, 25th and 75th
percentiles of the loss distribution (gray dashed curves).
}
    \label{fig: train loss}
\end{figure}

Training each neural network for $10^6$ epochs may seem %initially appear 
excessive; however, even over such an extended period, the loss function exhibits rapid changes within a small number of 
optimization steps %, 
occurring unpredictably at any stage, as illustrated in Fig.~\ref{fig: train loss} (left panel)
where we present evolutions of two different networks during the training on the same
datasets %data sets. 
%which include the NCSM results for the $^{6}$Li ground state energy obtained with different~$\hbar\Omega$ values in the model spaces $N_{\max}=2$,
%$N_{\max}=4$, ...\,, $N_{\max}=N_{\max}^{u}$ 
with $N_{\max}^{u}=18$.
Moreover, the right panel of Fig.~\ref{fig: train loss} 
demonstrates that the ensemble as a whole continues to improve %, 
with the median loss function
value (green solid curve) % (depicted by the green line) 
steadily decreasing throughout the training process. The blue dash-dotted
and  red dotted curves % lines 
represent the minimal and the
\mbox{maximal} loss function
values, respectively, while the gray dashed curves are the plots of % lines denote 
the 25th and 75th percentiles of the loss distribution. 
The loss function values presented in Fig.~\ref{fig: train loss}
%To enhance clarity, loss function values 
were taken only at the end of each training cycle. % {\color{red}to enhance clarity}. %when plotting the graph. 
These observations suggest that an
early stopping may not be an optimal strategy in this 
problem. %context.
%
%\begin{figure}[t]
%    \centering    
%    \includegraphics[width=0.49\linewidth]{id=(3,205).eps}
%    \includegraphics[width=0.49\linewidth]{loss_plot_for_paper.eps}
%    
%    \caption{Loss functions %value during 
%training on the $^{6}$Li %\(^6\text{Li} (3^+,0)\) energy data 
%ground state energy datasets with \(N_{\max}^u=18\). Left panel: evolutions of two
%different networks during the training; right panel: median loss function of the network ensemble  (green solid curve),  minimal (blue dash-dotted curve)
%and  maximal (red dotted curve) loss function values, 25th and 75th
%percentiles of the loss distribution (gray dashed curves).
%}
%    \label{fig: train loss}
%\end{figure}

This  behavior may be  attributed 
 to the cyclical learning rate schedule.
This behavior may be  
% {\color{blue}also be} %be also 
related also %attributed 
to the phenomenon of grokking \cite{grokking2022}, where an ensemble with the
increasing number of neural networks undergoes %in the ensemble undergo 
a phase transition, shifting to significantly lower loss values. 
%Also this behavior may be  addressed  to the cyclical learning rate schedule. 
As a result, while individual networks may experience abrupt improvements, the ensemble as a whole exhibits a relatively smooth and continuous learning trajectory. However, an
additional research is required to explicitly demonstrate the effect of grokking.

The set of hyperparameters outlined in this section (but not limited to) controlling the %which control 
machine learning process has been designed and carefully tested in 
%was carried out while considering 
the described above benchmark %extrapolation 
problem of extrapolating % predicting 
the ground state energy of the $^2\text{H}$ nucleus supported by the Nijmegen~II $NN$ potential.

\subsection{Prediction processing}
\label{Method: prediction processing}

Not all trained neural networks produce reasonable predictions, thus it is needed %making it essential 
to establish selection criteria for networks in 
the ensemble. The selection process consists of three main stages~\cite{Mazur2024,sharypov2024machine}: % firstly, 
constructing a prediction array 
including results from %for 
different model spaces, %secondly, 
applying filtering criteria to retain only reliable networks, and %thirdly, 
filtering the outliers.

\paragraph{Prediction array construction}
\label{Prediction array construction}
The %A 
prediction array is obtained by generating the results %generated  
for model spaces %$N_{\max} = N_{\max}^{u}$, 
$N_{\max}^{u} + 2$, $N_{\max}^{u} + 4$,  ...\,, %\dots, 
$N_{\max}^{f}$ %, 
and the
range of $ \hbar \Omega $ values matching %matches 
that of the training dataset. We use \(N_{\max}^{f} = 300\) as the upper limit for the model space
predictions.

\paragraph{Selection criteria}
\label{Selection Criteria}

To ensure the reliability and stability of the ensemble, we apply the following selection criteria:
\begin{itemize}

\item  ``Soft'' %"Soft" 
Variational Principle (for energies only): The predictions %Predictions 
obtained with increasing $N_{\max}$ at the same $\hbar\Omega$ values
may exhibit minor violations of %deviations from 
the variational principle due to numerical errors %approximation effects. 
which are not exceeding 0.1~keV.
%However, the total violation at \(N_{\max}^{f} = 300\) must not exceed 0.1~keV.

\item $\hbar\Omega$ Independence (or weak dependence): %on $\hbar\Omega$: 
At  $N_{\max}^{f} = 300$, the
predictions should be stable with respect to the $\hbar\Omega$ variations. %in $\left(\hbar\Omega\right)$. 
We impose the maximal %a maximum 
allowable deviations %deviation 
%0.002~MeV 
{%\color{red}
of 2~keV} for energies %energy 
and of
0.002~fm for radii.

%\item Convergence Principle (for energies only): For each network in the ensemble, we define a critical %model space %size 
%$N_{\max}^{c}$  such that the network predictions for model spaces $N_{\max}\geq N_{\max}^{c}$ 
%fit %from which 
%the following two conditions: % hold:
%(1) For any %a 
%fixed model space $N_{\max}\geq N_{\max}^{c}$, the difference between the 
%predicted maximal and minimal energies %maximum and minimum energy values as a function of \(\hbar\Omega\)  %must 
%in the range of \(\hbar\Omega\) values included in the prediction array do 
%not exceed 0.02~MeV.
%(2)~The difference between the minimal %minimum 
%energy values in adjacent model spaces $N_{\max}$ and $N_{\max} + 2 $ do
%   $\left( N_{\max} \right)$ and $\left( N_{\max} + 2 \right)$ must 
%not exceed 0.001~MeV.

{%\color{blue}
\item Convergence Principle (for energies only): For each network in the ensemble, a critical model space %size 
$N_{\max}^{c}$ is defined such that the network predictions for this model space
fit %from which 
the following two conditions. %: % hold:
(1)~In the model space $N_{\max}=N_{\max}^{c}$,  %For a fixed model space $N_{\max}^{c}$,
 the difference between the 
predicted maximal and minimal energies %maximum and minimum energy values as a function of \(\hbar\Omega\)  must 
in the range of \(\hbar\Omega\) values included in the prediction array do 
not exceed 20~keV.
(2)~The difference between the predicted
minimal %minimum 
energy values in adjacent model spaces $N_{\max}=N_{\max}^c$ and $N_{\max}=N_{\max}^c + 2 $ do
%   $\left( N_{\max} \right)$ and $\left( N_{\max} + 2 \right)$ must 
not exceed 1~keV.
}

From the networks that meet these criteria, the 20\% with the highest
$N_{\max}^{c}$ values are discarded to ensure fast enough convergence behavior.

\item Loss-Based Filtering: The %To further refine the ensemble, the 
5\% of networks with the largest %highest 
 loss function values at the end of training are also discarded.

\end{itemize}

\paragraph{Outlier Filtering}
\label{Outlier Filtering}

Following the above 
initial selection procedure, an additional outlier filtering step is performed~\cite{sharypov2024machine}. 
For each neural network, predictions at $N_{\max} = N_{\max}^{f}$ are taken. In the case of energy predictions, the minimal %minimum 
value is determined 
for the range of $\hbar\Omega$ values. For the
radius predictions, the arithmetic mean is calculated for the range of $\hbar\Omega$.

This yields %in 
a distribution of extrapolated energy or radius values in the network ensemble. 
To remove outliers from this distribution, we apply the boxplot rule as %it 
described in Ref. \cite{wilcox2009basic}.

A neural network is excluded from further statistical analysis if it fails to meet any of the above selection criteria. This ensures that only models exhibiting stable and physically consistent predictions contribute to the final ensemble results, enhancing the reliability of the overall approach.

\paragraph{Final Estimation and Uncertainty Quantification} 
\label{Method: Final Estimation and Uncertainty Quantification}

For the filtered prediction array at $N_{\max}^f$, the median value is computed and serves as the final estimate of the observable. The interquartile range between %(with 
the 25th and 75th percentiles % indicated in the tables below) 
is used as a measure of the extrapolation %method's 
uncertainty.
% {\color{red}This uncertainty indirectly reflects the uncertainty in the trained parameters of the single employed machine learning model.}
%This uncertainty indirectly reflects the uncertainty in the trained parameters of the single employed machine learning model.

\section{Results}
\label{Results}  
%\lipsum[1-3]

%\colorbox{yellow}{HERE}

The \( ^6 \)Li nucleus is of particular interest since its ground state
energy and rms point-proton
%root-mean-square (rms) point proton 
radius \( r_p \) have been %were previously 
studied using the
machine learning %-based 
extrapolation in the pioneering work~\cite{Negoita2019}. Moreover, the   %both   Ref.~\cite{Negoita2019} 
%and our study use 
training datasets obtained in the %derived from 
NCSM calculations with the Daejeon16 $NN$ interaction 
were utilized in our study and in Ref.~\cite{Negoita2019} as well as in a recent investigation of Ref.~\cite{Knol2025} 
which benchmarks  two
%where a benchmarking of 
different approaches to the neural network training, %was performed.
one is the approach of   Ref.~\cite{Negoita2019} and the other is the approach developed in 
Refs.~\cite{KNOLL2023137781,WolfgruberPhysRevC.110.014327}. There are
%However, there are
key differences %exist 
between these %the two 
approaches: Ref.~\cite{Negoita2019} suggests training an ensemble of artificial neural networks for each observable in each particular nucleus
while the  Refs.~\cite{KNOLL2023137781,WolfgruberPhysRevC.110.014327} suggest an ensemble of
universal networks for a set of nuclei %trined 
trained on a particular observable in lightest nuclei. The approach of Ref.~\cite{Negoita2019} is close to ours,
however, there are essential differences
 including the neural network topology, selection criteria for trained networks, and several other methodological aspects. % discussed earlier.

\renewcommand{\arraystretch}{1.35}
\begin{table}[t!]
    \centering\small
    \begin{tabular}{|c|c|c|c|c|c|}    
    
    \hline \parbox{6ex}{Nucleus (state)} & \parbox{5ex}{Var. min.} & Extrap. B & This work & Other & Experim.\\
    
    \hline $^6\text{He}$ (g.\,s.) & $-29.377$ & $-29.41(1) %\pm 0.01
    $ & $-29.429^{+0.007}_{-0.005}$ & $-$ & $-29.269$\\% \cite{TILLEY20023}\\
    
    \hline $^6\text{Li}$ (g.\,s.) & $-31.977$ & $-32.007(9) %\pm 0.009
    $ & $-32.036(3)%^{+0.003}_{-0.003}
     $ & \parbox{21ex}{ $-32.061(4)$ ISU-a %\cite{Negoita2019} 
     $-32.036(20)$ ISU-b
    % $-32.011(4)$ 
     $-32.011^{+0.006}_{-0.014}$ TUDa}& 
     $-31.995$ \\ %\cite{TILLEY20023}\\
    
    \hline $^6\text{Li}$ ($3^+\!$,0) & $-30.072$ & $-30.102(12)%\pm 0.012
    $ & $-30.129^{+0.004}_{-0.003}$ & $-$ & $-29.809$\\
    
    \hline $^6\text{Li}$ ($0^+\!$,1) & $-28.481$ & $-28.507(4)% \pm 0.004
    $ & $-28.552^{+0.008}_{-0.005}$ & $- $& $-28.434$\\    
    
    \hline
    \end{tabular}
    \caption{    
    Extrapolation results for the energies (in MeV)
of 
the $^{6}$He and $^{6}$Li     ground states
and the lowest excited states in $^{6}$Li obtained using the %with 
largest training dataset ($N_{max}^u = 18$) in comparison with the variational minimum obtained in the NCSM calculations at
${N_{\max}=18}$, with phenomenological extrapolation~B~\cite{Maris2009},
with machine learning approaches ISU-a of Ref.~\cite{Negoita2019} and ISU-b~\cite{Knol2025} utilizing a smaller range of $\hbar\Omega$ values for training,
with extrapolation TUDa using universal ensemble of neural networks~\cite{Knol2025},
%of {\color{blue} Ref.~\cite{Negoita2019, Knol2025}}, 
and with experiment~\cite{TILLEY20023}. The symmetric uncertainties of the last presented digits
are given in parentheses  while asymmetric  uncertainties are given in super and subscripts.
%All energies %Energies 
%are given in MeV.~%units of megaelectronvolts (MeV)
}
    \label{tab: energy}
\end{table}
\renewcommand{\arraystretch}{1.0}

{%\color{blue}
The extrapolated energy values for the $^{6}$He ground state and a few lowest states in $^{6}$Li, along with their associated uncertainties, are summarized in Table~\ref{tab: energy}.}
For comparison, %we present
%in Table~\ref{tab: energy}
%{\color{blue}
we %also 
present in Table~\ref{tab: energy} %. this Table}
%also 
the results obtained by the phenomenological exponential extrapolation~B~\cite{Maris2009}, % and 
by the machine learning extrapolation approach of Ref.~\cite{Negoita2019} (ISU-a),
%{\color{blue}Ref.~\cite{Negoita2019, Knol2025}} 
by the same approach utilizing a smaller range of $\hbar\Omega$ values for the network
training (ISU-b)~\cite{Knol2025}, 
by the approach of    Refs.~\cite{KNOLL2023137781,WolfgruberPhysRevC.110.014327} utilizing the ensemble of
universal networks (TUDa)~\cite{Knol2025},
as well as the respective experimental data~\cite{TILLEY20023} and
variational minima in the NCSM calculations at $N_{\max}=18$, i.\,e., in the largest model
space used for the training of neural networks. 
Our energy predictions lie below those obtained by the %via the phenomenological exponential 
extrapolation~B clearly 
demonstrating inaccuracy of this phenomenological approach. %method B~\cite{Maris2013} but, 

Our previous studies \cite{Mazur2024,sharypov2024machine} showed that a proper selection of the training dataset is essential for 
obtaining reliable predictions of extrapolated observables. For the energy extrapolations, we limited the training dataset to %values of 
the \( \hbar\Omega \) values ranging from the variational minimum for each model space up to 
40~MeV. The ISU-a  predictions were obtained based on the NCSM data spanning the \( \hbar\Omega \) range from %of 
8 to 50 MeV while the ISU-b results  were obtained with the  same neural network model  trained on the data from the  \(\hbar \Omega \) range %of 
from 10 to 30 MeV.  As seen in Table~\ref{tab: energy},
the choice of the training dataset significantly impacts both the convergence behavior and the final predictions for the energies.
In the case of the $^{6}$Li ground state, our extrapolated energy lies above the ISU-a result and is consistent with the ISU-b estimation which, however,
is characterized by a much larger uncertainty.  Notably, the estimated ISU-a uncertainty interval 
does not overlap with those obtained by other machine learning approaches indicating that that one should be very accurate in applications of these
approaches to extrapolating nuclear observables. It is important to emphasize that our approach emploing a larger ensemble of trained neural networks and strict %stricter 
 selection criteria, yields predictions with higher statistical reliability and weaker dependence on \( N_{\max}^u \). 

\begin{figure}[t!]
    \centering
    \includegraphics[width=0.5\linewidth]{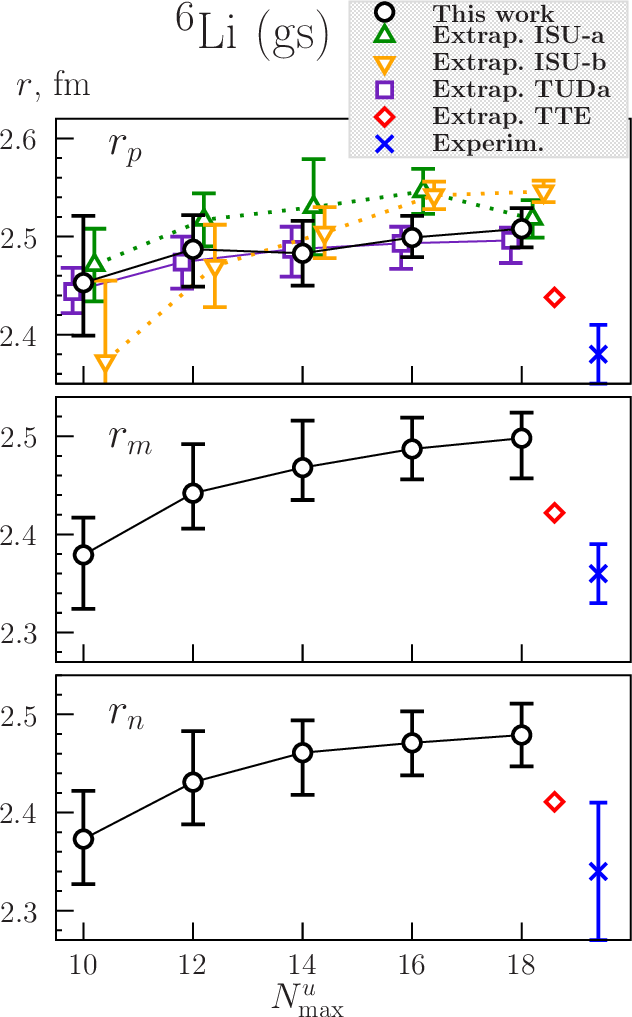}
    \caption{%Extrapolation results of point-proton ($r_p$), matter ($r_m$) and point-neutron ($r_n$) radii for the $^6\text{Li}$ in ground state with increase of training data set (denoted by $N_{max}^u$). Radii are given in units of femtometers (fm)
    Extrapolation results (black circles% with error bar
) for rms point radii $r_p$, $r_m$ and  $r_n$ 
in the  $^6$Li  ground state with increasing training data sets including the NCSM results obtained
with  $N_{\max}\leq N^u_{\max}$. For comparison: blue oblique crosses present experimental values~\cite{Tanihata2013}, green triangles,  orange
 triangles and violet diamonds present respectively %ANN 
machine learning
extrapolations ISU-a~\cite{Negoita2019},  % of 
%Ref.~\cite{Negoita2019}, orange triangles  present modified ANN extrapolations 
ISU-b~\cite{Knol2025} and %ISU (ISU-b,  taken from Ref.~\cite{Knol2025}), 
%violet square present ANN extrapolations 
TUDa~\cite{Knol2025}, % Ref.~\cite{Knol2025} 
and red diamonds present TTE extrapolations~\cite{Rodkin2023}. %of 
%Ref.~\cite{Rodkin2023}.
}
    \label{fig: 6Li}
\end{figure}

For the rms radii extrapolations. as follows from our  studies in Refs.~\cite{Mazur2024,sharypov2024machine}, 
the optimal training datasets include the NCSM results obtained with $\hbar\Omega $ values ranging from %10 
12.5 to 40~MeV. The convergences of the
extrapolated rms point-proton, point-neutron, and point-nucleon (matter) %matter 
radii of the \( ^6 \)Li ground state %, along with their convergence behavior 
as \( N_{\max}^u \) increases, are presented in 
Fig.~\ref{fig: 6Li}. 
The final results obtained  using the largest training dataset which includes all NCSM 
results from calculations in the model spaces $N_{\max}\leq N_{max}^u = 18$, can be found in
%and 
Table~\ref{tab: radii}. Our predictions exceed both experimental values and the phenomenological $2D$ twisted tape extrapolation (TTE)
%extrapolation 
results~\cite{Rodkin2023} based also on the NCSM calculations with Daejeon16. 

The importance of a proper  selection of the training
dataset is clearly seen in Fig.~\ref{fig: 6Li} where our results are compared with machine learning extrapolations ISU-a~\cite{Negoita2019} and 
ISU-b~\cite{Knol2025} which differ only by the range of $\hbar\Omega$ values included in the sets of the NCSM results for the training. The convergence
of the \mbox{ISU-b} version is clearly worse and the final ISU-a and ISU-b results at $N^{u}_{\max}=18$ only slightly overlap if the uncertainties are
taken into account. From our experience, it is better to exclude small $\hbar\Omega$ values in the training datasets for the radii extrapolations. Our
%However, unlike the energy extrapolations, our 
predicted rms point-proton radius \( r_p \) is in agreement with the results of 
machine learning approaches ISU-a and TUDa. % of Ref.~\cite{Negoita2019}.

\renewcommand{\arraystretch}{1.35}
\begin{table}[t]
    \centering\small
        \begin{tabular}{|c|c|c|c|c|c|}
        \hline  \parbox{7ex}{Nucleus (state)} & \parbox{4ex}{rms radii} & 
        \parbox{6ex}{Cross point} &This work & Experim. %\cite{Tanihata2013} 
    & Other %TTE \cite{Rodkin2023, PhysRevC.106.034305} 
    \\
        
        \hline \multirow{3}{*}{$^6\text{He} (g.s.)$} & $r_p$ & 1.86 & 
        %$1.926^{+0.018}_{-0.023}$
        $1.93(2)$ & $1.925(12)$ & \parbox{20ex}{ $1.871(6)$ TTE} \\
        
        \cline { 2 - 6 } & $r_m$ & 2.42 & %$2.504^{+0.031}_{-0.041}$ 
        $2.50^{+0.03}_{-0.04}$ & $2.50(5)$ &  \parbox{20ex}{ $2.430(6)$ TTE}  \\
        
        \cline { 2 - 6 } & $r_n$ & 2.64 &% $2.769^{+0.017}_{-0.044}$ 
        $2.77^{+0.02}_{-0.04}$ & $2.74(7)$ & \parbox{20ex}{ $2.663(3)$ TTE}   \\
        
        \hline \multirow{3}{*}{$^6\text{Li (g.s.)}$} & $r_p$ & 2.44 & 
       % $2.508^{+0.021}_{-0.019}$ 
        $2.51(2)$& $2.38(3)$ & \parbox{18ex}{ 2.411 \  TTE\hphantom{i}
    $2.518(19)$ ISU-a 
  $2.546(11)$ ISU-b   
   %$2.515^{+0.014}_{-0.030}$  
  $2.496^{+0.013}_{-0.023}$ TUDa}\\

        \cline { 2 - 6 } & $r_m$ & 2.44 &% $2.498^{+0.026}_{-0.041}$ 
        $2.50^{+0.03}_{-0.04}$ & $2.36(3)$ & \parbox{20ex}{ 2.422 TTE}   \\
        
        \cline { 2 - 6 } & $r_n$ & 2.43 & %$2.479^{+0.032}_{-0.032}$ 
        $2.48(3)$& $2.34(7)$ & \parbox{20ex}{ 2.438 TTE}  \\

        \hline \multirow{3}{*}{$^6\text{Li} (3^+,0)$}  & $r_p$ & 2.38 & 
        %$2.450^{+0.043}_{-0.057}$ 
        $2.45^{+0.04}_{-0.06}$ & - & - \\
        
        \cline { 2 - 6 } & $r_m$ & 2.26 & %$2.421^{+0.088}_{-0.088}$ 
        $2.42(9)$ & - & - \\
        
        \cline { 2 - 6 } & $r_n$ & 2.30 & %$2.441^{+0.072}_{-0.094}$
        $2.44^{+0.07}_{-0.09}$ & - & - \\

        \hline \multirow{3}{*}{$^6\text{Li} (0^+,1)$}  & $r_p$ & 2.48 & 
        %$2.664^{+0.064}_{-0.088}$ 
        $2.66^{+0.06}_{-0.09}$ & - & - \\
        
        \cline { 2 - 6 } & $r_m$ & 2.42 & $2.61(5)$ %$2.611^{+0.050}_{-0.052}$ 
        & - & - \\
        
        \cline { 2 - 6 } & $r_n$ & 2.40 & $2.59(5)$%2.50^{+0.048}_{-0.053}$ 
        & - & - \\

        %\hline \multirow{3}{*}{$^6\text{Be}$} & $r_p$ & $3.162\binom{+0.085}{-0.094}$ & - & - \\
        
        %\cline { 2 - 5 } & $r_m$ & $2.757\binom{+0.086}{-0.050}$ & - & - \\
        
        %\cline { 2 - 5 } & $r_n$ & $2.077\binom{+0.037}{-0.038}$ & - & - \\
        
        \hline
        
    \end{tabular}    
    \caption{Extrapolated %Extrapolation results of 
point-proton $r_p$, point-nucleon $r_m$, %($r_p$), matter ($r_m$) 
and point-neutron $r_n$ % ($r_n$) 
radii (in fm) in  the $^6$He and $^{6}$Li ground
states and in the lowest excited states in  $^{6}$Li
obtained using %with 
the largest training dataset ($N_{max}^u = 18$) %Radii are given in units of femtometers (fm)
in comparison with experimental data~\cite{Tanihata2013}, TTE extrapolations from Ref.~\cite{Rodkin2023} for $^{6}$Li and from 
Ref.~\cite{PhysRevC.106.034305} for  $^6$He and machine learning extrapolations ISU-a~\cite{Negoita2019},  ISU-b~\cite{Knol2025} and 
TUDa~\cite{Knol2025}. The symmetric uncertainties of the last presented digits are given
in parentheses while asymmetric uncertainties are given in super and subscripts.
}
    \label{tab: radii}
\end{table}
\renewcommand{\arraystretch}{1.0}

As compared  to the ground state, in the case of %For 
the first excited state \( (3^+,0) \) of the \( ^6\text{Li} \) nucleus, the uncertainties of %in 
the extrapolated values of the rms point
radii \( r_p \), \( r_m \), and \( r_n \) are significantly larger (see Table~\ref{tab: radii}). %compared to the ground state. 
However, the overall convergence remains reasonably good. Similar to the ground state, the values of \( r_p \), \( r_m \), and \( r_n \) are close to each other. The central values of all
these radii %, though all three 
are slightly smaller  the %{\color{blue}than the} 
respective values %than 
in the ground state but overlap with them if the extrapolation uncertainties are taken into account.  

\begin{figure}[t]
    \centering
    \includegraphics[width=0.5\linewidth]{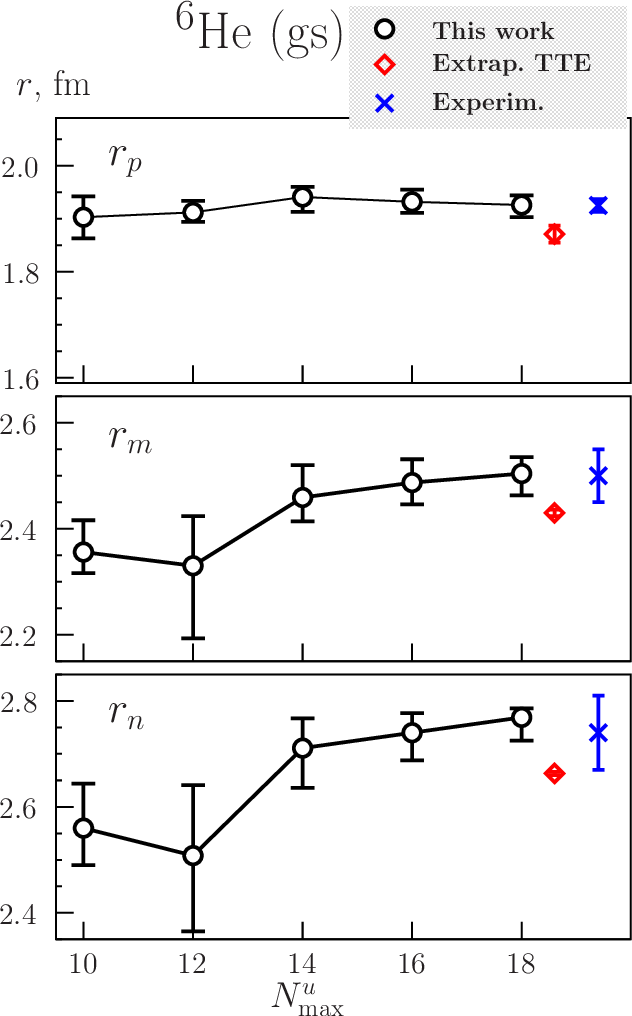}
    \caption{%Extrapolation results of point-proton ($r_p$), matter ($r_m$) and point-neutron ($r_n$) radii for the $^6\text{He}$ in ground state with increase of training data set (denoted by $N_{max}^u$). Radii are given in units of femtometers (fm)
    Extrapolation results (black circles) for rms point radii $r_p$, $r_m$ and  $r_n$ in the $^6$He 
ground state with increasing training data sets including the NCSM results obtained
with  $N_{\max}\leq N^u_{\max}$. For comparison: blue oblique crosses present experimental values~\cite{Tanihata2013} and %,  
red diamonds present TTE extrapolations~\cite{PhysRevC.106.034305}.
% of Ref.~\cite{PhysRevC.106.034305}.
}
    \label{fig: 6He}
\end{figure}

The second excited state \( (0^+,1) \) of \( ^6\text{Li} \) and the ground states \( (0^+,1) \) of \( ^6\text{He} \) and \( ^6\text{Be} \) nuclei
 form an isospin triplet. Comparing the energies and rms radii of these states provides valuable insights into their structural similarities and differences. 
The convergence of the $^{6}$He rms point radii with \( N_{\max}^u \) is presented in
Fig.~\ref{fig: 6He}.
%the final result obtained with the largest training dataset can be found in Table~\ref{tab: radii}. 

Unlike the \( (0^+,1) \) state in
\( ^6\text{Li} \), the rms point-proton radius \( r_p \) of the \( ^6\text{He} \) nucleus, 
as seen in Table~\ref{tab: radii}, %in its ground state 
is much
smaller than its %the 
point-neutron radius \( r_n \) revealing the two-neutron halo structure of this nucleus. 
The rms point-nucleon %matter 
radius \( r_m \) lies %falls 
between these two values satisfying %, and 
with a
good accuracy  %, 
the well-known relation% is satisfied:
\begin{equation}
    A r_m^2 = Z r_p^2 + N r_n^2 .
\end{equation} 

A comparison of predicted rms point radii
with the
experimental data and the results of the phenomenological $2D$ TTE
extrapolation~% method 
\cite{Rodkin2023, PhysRevC.106.034305} is presented in Table~\ref{tab: radii}. Notably, our 
$^{6}$He
results % {\color{green}nicely} 
agree with the experiment for radii %experimental values within the given uncertainties, 
while the predicted $^{6}$He ground state %ground-state 
energy (see Table~\ref{tab: energy})
lies {%\color{green} 
slightly} %significantly 
below the experimental value.

\section{Discussion}
\label{Discussion}

Using our modification~\cite{Mazur2024,sharypov2024machine} of
the %ensemble neural network 
machine learning method proposed in Ref.~\cite{Negoita2019}, 
%and modified in our research \cite{Mazur2024,sharypov2024machine}, 
we perform   %ed 
extrapolations to the
 infinitely large model space of the  %s for the ground-state 
 NCSM results for
 energies \(E\) and rms %root-mean-square (rms) 
point-proton \(r_p\), point-neutron \(r_n\), and point-nucleon (matter) \(r_m\) radii of the 
\(^{6}\)He nucleus in the ground state %\(^{6}\)Li, \(^{6}\)He, and \(^{6}\)Be, as well as of the  %for the first two excited states of \(^{6}\)Li.
and  in the three lowest states of the \(^{6}\)Li nucleus. The NCSM calculations were performed using the Daejeon16 $NN$ interaction in
model spaces up to $N_{\max} = 18$.
%
%The neural networks were trained on datasets consisting of pairs \((N_{\max}, \hbar\Omega)\) and \(E\) (or \(r_p\), \(r_n\), or \(r_m\)) computed within the No-Core Shell Model using the Daejeon16 NN interaction in model spaces up to \(N_{\max} = 18\).  
%
The training datasets for the neural networks consist %consisted 
of the NCSM parameters  \(N_{\max}\) and \(\hbar\Omega\)  %parameter pairs \((N_{\max}, \hbar\Omega)\) 
alongside with corresponding computed values of \(E\) (or \(r_p\), \(r_n\), or \(r_m\)).
% within the No-Core Shell Model, utilizing the Daejeon16 $NN$ interaction in model spaces up to Nmax = 18.

The key innovation of  our  method is the training of an ensemble of artificial neural networks based on a preprocessing of training data, filtering trained networks according to rigorous selection criteria, and performing statistical post-processing of the accepted predictions. The extrapolation algorithm is implemented using Python with open-source machine learning libraries, including Keras \cite{chollet2015keras}, TensorFlow \cite{tensorflow2015-whitepaper}, and TensorFlow Addons \cite{TensorFlowAddons}.

The predictions for the %predicted 
ground state %ground-state 
energies of \(^{6}\)Li and \(^{6}\)He exhibit a
good convergence. By extrapolations utilizing %In calculations using 
the largest training dataset  including the NCSM results in model spaces up to $ N_{\max}^{u} = 18$ %(\(N_{\max}^{u} = 18\)), 
we obtain \(E(^{6}\text{Li}) = -32.036 \pm 0.003\) MeV and \(E(^{6}\text{He}) = -29.429_{-0.005}^{+0.007}\) MeV. The prediction uncertainties are comparable to the numerical accuracy of NCSM calculations (1 keV) and do not exceed the uncertainties of the machine learning %-based 
extrapolations ISU-a, ISU-b and TUDa. % reported in \cite{Negoita2019}.
 However, as it was shown in Refs.~\cite{Mazur2024,sharypov2024machine}, 
 at all values of \(N_{\max}^{u}\), our extrapolation
 results lie systematically higher than those of  ISU-a %\cite{Negoita2019} 
 including the uncertainties. %when considering error margins. 
 This suggests that different machine learning %-based 
 extrapolation approaches may lead to systematically different results. 
It is interesting that our prediction for the $^{6}$Li ground state energy  is consistent with the ISU-b prediction which uncertainty is, however, 
nearly an order of magnitude larger.
 Notably, due to the larger number of trained and selected neural networks in our study, the statistical reliability of our predictions is higher than those of ISU-a and ISU-b. % in \cite{Negoita2019}. 
The convergence trends of the TUDa approach for the energies is similar to ours 
but the final TUDa result suggests less binding of the $^{6}$Li nucleus and  does not overlap with ours accounting for the uncertainties.

%{\color{violet}
%It is interesting to note the strong agreement between the predictions of our approach and the TUDa model for both the energy and the rms point-proton radius, as well as the nearly identical convergence curves (see Fig.~\ref{fig: 6Li}). The TUDa model was trained on the same dataset as ISU-b.
%}

The prediction uncertainty for rms point-proton, point-neutron and point-nucleon (matter)  %neutron, and  matter 
radii is slightly larger, approximately 1\%, yet the overall convergence remains satisfactory. Our results for the radii of \(^{6}\)He are in excellent %good 
agreement with experimental data \cite{Tanihata2013}, while for \(^{6}\)Li, our predictions slightly exceed the experimental values. In general, our extrapolated rms
radii are a bit larger than  those %consistent with the results 
obtained using a two-dimensional phenomenological exponential extrapolation TTE %(TTE) approach applied to 
in the case of \(^{6}\)He~\cite{PhysRevC.106.034305} and \(^{6}\)Li~\cite{Rodkin2023} nuclei. %, based on NCSM calculations with the same Daejeon16 $NN$ interaction. 
The rms point-proton $^{6}$Li radii %radius 
predicted using the machine learning methods ISU-a and TUDa are close to ours while the ISU-b prediction lies above the results of all other extrapolations
and the experiment.
%in \cite{Negoita2019} is systematically larger than both our result and the experimental value.  
It is important to note the consistency of the predictions and
nearly identical convergence trends (see Fig.~\ref{fig: 6Li})
%he strong agreement between the predictions 
of our approach and that of the TUDa model for the rms point-proton radius in $^{6}$Li contrary to the ISU-a and ISU-b convergence patterns.
% , as well as the nearly identical convergence curves (see Fig.~\ref{fig: 6Li}). The TUDa model was trained on the same dataset as ISU-b.

%The key innovation in our modified method lies in training an ensemble of artificial neural networks based on a preprocessing of training data, filtering trained networks according to rigorous selection criteria, and performing statistical post-processing of the accepted predictions. The extrapolation algorithm is implemented with Python using open-source machine learning libraries, including Keras \cite{chollet2015keras}, TensorFlow \cite{tensorflow2015-whitepaper}, and TensorFlow Addons \cite{TensorFlowAddons}.  

Overall, our results indicate that machine learning offers a promising tool for addressing longstanding challenges in nuclear many-body theory.
We anticipate that the proposed extrapolation method is rather general and %it 
can be applied to other nuclear observables, such as quadrupole moments, electromagnetic transition probabilities, etc. %and beyond.

\section{Acknowledgements}
\label{Acknowledgements}
This work was supported by the Ministry of Science and Higher Education of the Russian Federation (project No. FEME-2024-0005)

\bibliographystyle{elsarticle-num} 
%\addbibresource{lib.bib}
\bibliography{lib.bib}




% %% For numbered reference style

%   \textit{\LaTeX: a document preparation system},
%   Addison Wesley, Massachusetts,

% \end{thebibliography}
\nocite{*}
%\printbibliography
\end{document}